\documentclass[11pt]{article}
\usepackage{cite}
\usepackage{tikz}
\usetikzlibrary{arrows,shapes}
\usetikzlibrary{trees}
\usetikzlibrary{matrix,arrows} 				
\usetikzlibrary{positioning}				
\usetikzlibrary{calc,through}				
\usetikzlibrary{decorations.pathreplacing}  
\usepackage{pgffor}							

\usetikzlibrary{decorations.pathmorphing}	
\usetikzlibrary{decorations.markings}
\tikzset{
	vector/.style={decorate, decoration={snake}, draw},
	provector/.style={decorate, decoration={snake,amplitude=2.5pt}, draw},
	antivector/.style={decorate, decoration={snake,amplitude=-2.5pt}, draw},
	fermion/.style={draw=black, postaction={decorate},
		decoration={markings,mark=at position .55 with {\arrow[draw=black]{>}}}},
	fermionbar/.style={draw=black, postaction={decorate},
		decoration={markings,mark=at position .55 with {\arrow[draw=black]{<}}}},
	fermionnoarrow/.style={draw=black},
	gluon/.style={decorate, draw=black,
		decoration={coil,amplitude=4pt, segment length=5pt}},
	scalar/.style={dashed,draw=black, postaction={decorate},
		decoration={markings,mark=at position .55 with {\arrow[draw=black]{>}}}},
	scalarbar/.style={dashed,draw=black, postaction={decorate},
		decoration={markings,mark=at position .55 with {\arrow[draw=black]{<}}}},
	scalarnoarrow/.style={dashed,draw=black},
	electron/.style={draw=black, postaction={decorate},
		decoration={markings,mark=at position .55 with {\arrow[draw=black]{>}}}},
	bigvector/.style={decorate, decoration={snake,amplitude=4pt}, draw},
}

\tikzstyle{block} = [draw, rectangle, 
minimum height=3em, minimum width=6em]

\usepackage{amsmath,amsfonts,amssymb}
\usepackage{graphicx,rotate,multicol,wrapfig}
\usepackage{color}
\usepackage{epsfig,epsf}
\usepackage{bm}
\usepackage{dcolumn,subfig}
\usepackage{array}

\usepackage{relsize}
\def\Babar{{\mbox{\slshape B\kern-0.1em{\smaller A}\kern-0.1em B\kern-0.1em{\smaller A\kern-0.2em R}}}}

\usepackage[bookmarks, breaklinks, colorlinks,urlcolor=magenta, citecolor=red, 
linkcolor=blue]{hyperref}

 \usepackage{color}

 \usepackage[normalem]{ulem}

 \definecolor{darkgreen}{cmyk}{1,0,1,0.4}
 
 \definecolor{pink}{cmyk}{0.4,1,0.3,0}

\def\com2#1{\textcolor{red}{\it{#1}}}

\def\bar {\overline}

\def\cs {\cal S}
\def\cat {\cal T}
\def\cu {\cal U}

\def\beq{\begin{equation}}
\def\eeq{\end{equation}}
\def\bea{\begin{eqnarray}}
\def\eea{\end{eqnarray}}
\def\barr{\begin{array}}
\def\earr{\end{array}}

\def\gev{\ensuremath{\mathrm{Ge\kern -0.1em V}}}

\begin{document}

\renewcommand*{\thefootnote}{\fnsymbol{footnote}}

\begin{center}
 {\Large\bf{Double Higgs boson production as an exclusive probe for a
 sequential fourth generation with wrong-sign Yukawa couplings}}\\[5mm]
Md.\ Raju$^{a,}$\footnote{mdraju@gmail.com}, 
Jyoti Prasad Saha$^{a,}$\footnote{jyotiprasadsaha@gmail.com},
Dipankar Das$^{b,}$\footnote{d.das@iiti.ac.in}, and
Anirban Kundu$^{c,}$\footnote{anirban.kundu.cu@gmail.com}\\[3mm]
{\small\em $^a$Department of Physics, University of Kalyani, Nadia 741235, India} \\ 
{\small\em $^b$Discipline of Physics, Indian Institute of Technology Indore, 
Khandwa Road, Simrol, Indore 453552, India}\\
{\small\em $^c$Department of Physics, University of Calcutta, 92
Acharya Prafulla Chandra Road, Kolkata 700009, India}

 \end{center}


\begin{abstract}
It has been shown that the data from the Large Hadron Collider (LHC) does not rule out a chiral sequential fourth 
generation of fermions that obtain their masses through an identical mechanism as the other three generations do.
However, this is possible only if the scalar sector of the Standard Model is suitably enhanced, 
like embedding it in a type-II two-Higgs doublet model. 
In this article, we try to show that double Higgs production (DHP) can unveil the existence of such a hidden 
fourth generation in a very efficient way. While the DHP cross-section in the SM is quite small, it is significantly 
enhanced with a fourth generation. We perform a detailed analysis of the dependence of the DHP cross-section 
on the model parameters, and show that either a positive signal of DHP is seen in
the early next run of the LHC, or the model is ruled out.

\end{abstract}



\setcounter{footnote}{0}
\renewcommand*{\thefootnote}{\arabic{footnote}}

\section{Introduction}

More than seven years after the discovery of the Higgs boson at the Large Hadron Collider (LHC) 
\cite{Aad:2012tfa, Chatrchyan:2012xdj}, the data have reached such a level of precision as to vindicate the 
Standard Model (SM) \cite{Khachatryan:2016vau} and to rule out a number of theories beyond the SM.  
The SM extended by a chiral fourth fermion generation
(SM4)\cite{Kribs:2007nz,Denner:2011vt,Eberhardt:2012gv,Djouadi:2012ae,Vysotsky:2013gfa,Kuflik:2012ai,Lenz:2013iha} constitutes such an example, 
as the quantum effects of SM4 on the production and decay of the Higgs boson do not decouple even in the 
heavy mass limit\cite{Gunion:1989we}. Especially, the production rate through 
	gluon-gluon fusion, $gg\to h$, shoots up way beyond 
	the experimental data.

However, it was recently shown in Ref.\ \cite{Das:2017mnu} that if the fourth generation is augmented by an extra Higgs doublet,
it is possible to hide such quantum effects completely, albeit in a certain case known as the wrong-sign 
limit\cite{Fontes:2014tga,Ferreira:2014dya,Biswas:2015zgk,Ferreira:2014naa}. The
ambiguity in the determination of the sign of the down-type Yukawa couplings plays a crucial role in arranging
cancellation among the amplitudes mediated by the fourth generation fermions in the Higgs production and 
decay processes\footnote{The sign of the top Yukawa coupling with respect to the $WWh$ coupling 
is fixed from the diphoton decay width of the Higgs boson.}. 
We achieve this conspiracy of Yukawa couplings in the framework of a type-II 2HDM.
We would like to draw the attention of the readers to a couple of points here. First, there can be other extensions of 
the SM that can accommodate extra sequential generations of fermions in a similar way, the type-II two-Higgs doublet 
model (2HDM) is just an example. Second, such an arrangement
saves the fourth generation only from the Higgs data. Other constraints and issues, like 
those coming from the oblique parameters,
the stability of the potential, the scattering unitarity, and the inherent non-perturbative nature of the fourth generation 
Yukawa couplings, must have to be taken care of separately, possibly by the introduction of other degrees of 
freedom and new dynamics associated with them.

When one talks about a second Higgs doublet, the data on the production and decay of the 125 GeV scalar resonance 
must be taken into account. In the framework of a 2HDM, which we will take to be type-II for our discussion, this means
that we must be close to the alignment limit
\cite{Gunion:2002zf,Carena:2013ooa,Dev:2014yca,Das:2015mwa,Bhattacharyya:2015nca}, {\it i.e.}, 
the lighter CP-even neutral scalar $h$ must be SM-like in its tree-level
couplings to fermions and gauge bosons, modulo the phase of some of the couplings. The alignment limit and the 
wrong-sign limit are not the same, but they are close to each other when $\tan\beta \, (=v_2/v_1)$, the ratio of the 
vacuum expectation values (VEV) of the two Higgs doublets, is large and is related to the other parameters in
a specific way.

Assuming that such a model -- the SM extended by a heavy chiral fourth generation of leptons and quarks, and another
Higgs doublet of type-II variety, the complete package of which we will call xSM4 -- exists, and somehow satisfies all 
the constraints mentioned in the previous paragraphs, one would like to ask the pertinent question of the possible direct 
signatures of the model. There can be indirect signals through flavour physics, coming from the mixing of the fourth 
generation with the other three, but that involves parameters that are {\em a priori} unknown. Similarly, direct 
production of heavy fermions also depends on unknown quantities like the masses of those fermions. 
The model can be so tuned as to make all such effects vanish.

There is, however, an interesting way to unveil the fourth generation, 
which we would like to focus on in this paper. This is the double Higgs production (DHP), $pp \to hh + X$. 
The DHP is the most important process to directly measure the Higgs self-coupling, and in context of the LHC, this has been
widely discussed in the literature
\cite{Baur:2003gp,Baur:2002qd,Baur:2002rb}.
DHP can proceed through various subprocesses: gluon-gluon fusion (ggF), vector boson fusion (VBF), 
Higgsstrahlung, or bremsstrahlung from the top; but ggF is by far the dominant process\cite{DiMicco:2019ngk,Plehn:1996wb,Frederix:2014hta}, contributing more than 
90\%. At $\sqrt{s}=13$ TeV
at the LHC, the total DHP cross-section in the SM is 34.45 fb, of which 31.05 fb comes from ggF
\cite{DiMicco:2019ngk}. At $\sqrt{s}=14$ TeV, the ggF share is 36.69 fb out of a total of 40.71 fb.
In the ggF channel, there are two types of Feynman diagrams in the SM; 
one is represented by a top quark box, and the 
other by a top quark triangle and a triple Higgs interaction. The DHP rate in the SM
is suppressed because of a destructive interference between these two amplitudes. On the other hand,
in xSM4, the new quarks play a pivotal role (as well as the new scalars), and the 
DHP cross-section may receive an order-of-magnitude enhancement compared to the SM one, thanks to the large 
Yukawa couplings of the heavy fourth generation. We explain the mechanism in more detail in the next Section.    

We perform the analysis of DHP within the framework of xSM4 for $\sqrt{s}=14$ TeV, and show that 
(i) there should be a substantial enhancement of the DHP cross-section,
 compared to that in the SM,  which should
either result in a positive signal very soon, or rule the model out, and (ii) on top of the enhancement due
to the fourth generation of fermions, 
there is a further contribution from the heavy neutral Higgs of 2HDM, the effect of which results in a severe 
constraint on the parameter space of 2HDM associated with xSM4.

At this point, one must point out the inherent limitations of such a study. Heavy chiral fermions that get their
masses through Yukawa couplings, naturally engender those couplings to be large and possibly nonperturbative. 
In this limit, the one-loop calculations have to be taken with a pinch of salt, as the higher-loop electroweak corrections
may be even more significant \cite{Djouadi:1997rj,Denner:2011vt}.
However, the masses of the fourth generation fermions can be tuned to arrange
a cancellation\cite{Djouadi:1997rj} and consequently such effects can be diluted.
In any case, such contributions should, in principle, raise the DHP cross-section, if
the higher-order terms add constructively to the leading order ones. In other words, our bounds should turn out 
to be even stronger, or, in an extreme case, the model may already be ruled out. The ideal procedure would 
have been to use an effective theory, involving a $G^{a\mu\nu}G^a_{\mu\nu}\Phi^\dag\Phi$ operator ({\em i.e.}, 
involving two gluon and two Higgs fields), and integrating out the heavy fermion and scalar fields. However, 
this brings in an undetermined Wilson coefficient, whose value must be ascertained by matching with the 
full theory. Thus, one has to evaluate the full ultraviolet-complete theory, as best as possible.

The paper is organized as follows. In Section 2, we briefly recapitulate the DHP mechanism in the SM, 
and also give a brief introduction to xSM4. The DHP in xSM4 is analyzed in Section 3, with the 
results are shown and discussed in Section 4. In the last Section, we summarize our findings and conclude.

\section{Theory prelude}
\subsection{Double Higgs production in the SM} \label{s:dhpsm}
\begin{center}
\begin{figure}[htbp!]
 \includegraphics[width=17cm]{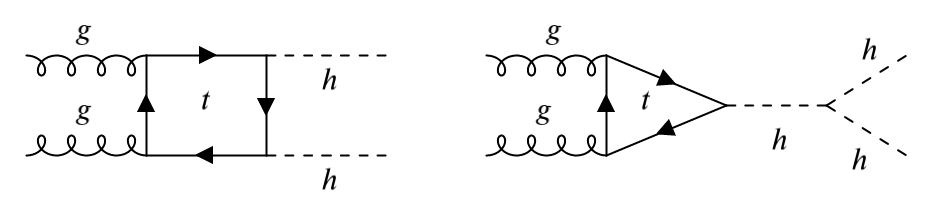}
 	\caption{Feynman diagrams for double Higgs Production in the SM.} \label{hhsm}
\label{f:SMfeynman}
\end{figure}
\end{center}

The Feynman diagrams relevant for DHP in the SM ggF channel 
are shown in  Fig.~\ref{hhsm} (for all the diagrams we refer the reader to the recent review, 
Ref.\ \cite{DiMicco:2019ngk}). The subdominant contributions are neglected for our discussion. 

The first diagram depicts an amplitude that proceeds through a quark box; this will be called a 
{\em Box} (or $\Box$) diagram. The second diagram similarly depicts an amplitude proceeding through a quark
triangle and subsequent triple-Higgs interaction; this is called a {\em Delta} (or $\Delta$) diagram. Note that 
there are many $\Box$ and $\Delta$ diagrams for different quarks; however, in the SM, one may neglect 
all other quarks except the top.

An interesting aspect of DHP in the SM is the destructive interference between the Box and the Delta 
amplitudes. This results in an extremely small cross section, of about 36.69~fb
at the 14 TeV LHC \cite{DiMicco:2019ngk}. 
Searches for both resonant and non-resonant Higgs pair production have been performed in various 
channels by both the ATLAS and CMS 
experiments~\cite{Aad:2015xja,ATLAS:2016ixk,ATLAS:2016qmt,CMS:2016tlj,
CMS:2016pwo,CMS:2016vpz,
CMS:2016rec,CMS:2017orf,CMS:2017ums,CMS:2016ymn,TheATLAScollaboration:2016ibb}.


Let us display the relevant expressions for the DHP differential cross-section via the ggF
channel. The process under consideration 
is $g(p_1)\, g(p_2) \, \to \, h(p_3)\, h(p_4)$, where $p_1$ and $p_2$ are the incoming momenta while 
$p_3$ and $p_4$ are the outgoing momenta, so that the Mandelstam variables are defined as
\begin{equation}
s = (p_1+p_2)^2\,,\ \ \ t = (p_1-p_3)^2\,,\ \ \ u = (p_1-p_4)^2\,.
\end{equation} 
To simplify the expressions, let us define some dimensionless quantities as follows:
\begin{equation}
{\cs} =\frac{s}{m^2_{t}}\,,\ \   {\cat}=\frac{t}{m^2_{t}}\,,\ \ {\cu}=\frac{u}{m^2_{t}}\,,\ \ 
\rho=\frac{m_h^2}{m_t^2}\,,\ \  
{\cat}_1={\cat}-\rho\,,\ \  {\cu}_1={\cu}-\rho\,, 
\end{equation}
where $m_t$ is the pole mass of the top quark and $m_h$ is the Higgs boson mass.
The spin and colour averaged differential cross-section for DHP is given by \cite{Glover:1987nx}
 \begin{equation}
\frac{d\sigma_{gg\to hh}}{dt} = 
\frac{G^2_{F}\alpha^2_{s}} {256\, (2\pi)^3}\, \left[\left\vert \left( C_{\Delta}\, F_{\Delta} + C_{\Box}\, F_{\Box}\right)
\right\vert^2 + \left\vert C_{\Box}\, G_{\Box}\right\vert^2 \right]\,,
\end{equation}
where $G_F$ and $\alpha_s$ $(\equiv g_s^2/ 4\pi)$ are the Fermi coupling and the strong coupling 
constants respectively. The expressions for the $F$ and $G$ terms as well as the coefficients 
$C_\Box$ and $C_\Delta$ are given below.
\begin{eqnarray}
C_{ij} &=&\int\frac{d^4q}{i\pi^2}\, \frac{1} { \left(q^2-m^2_{t}\right) \, \left[ \left( q+ p_{i}\right)^2 - 
m^2_{t}\right]\, \left[ \left( q + p_{i}+p_{j}\right)^2 - m^2_{t}\right]} \,,  \\
D_{ijk} &=& \int\frac{d^4q}{i\pi^2}\, \frac{1}
{\left(q^2-m^2_{t}\right)\, \left[ (q+ p_{i})^2-m^2_{t}\right]\, \left[ (q + p_{i} + p_{j})^2-m^2_{t}\right]\,
\left[ (q + p_{i} +p_{j} + p_{k})^2-m^2_{t}\right]}\,, \\
F_{\Delta} &=& \frac{2}{{\cs}}\, \left\{2+(4-{\cs})m^2_{t}C_{12}\right\}\,, \\
F_{\Box} &=& 
\frac{1}{{\cs}^2}\, \left\{ 4{\cs} + 8m^2_t\, {\cs} C_{12} - 2 m_t^4\,{\cs} ({\cs}+2\rho-8)\, (D_{123} + D_{213} +D_{132}) 
\right. \nonumber \\ 
&& \left. + 2 m_t^2 (\rho-4)\, \left[ {\cat}_1(C_{13}+C_{24})+{\cu}_1(C_{23}+C_{14})-m_t^2 \, ({\cat}{\cu}
-\rho^2)\, D_{132}\right]\right\}\,, 
\end{eqnarray}
\begin{eqnarray}
G_\Box &=& 
\frac{1}{{\cs}({\cat}{\cu}-\rho^2)}\, \left\{ m_t^2\, ({\cat}^2+\rho^2-8{\cat})\, \left[{\cs}C_{12}+{\cat}_1
(C_{13}+C_{24})- m^2_t\, {\cs}{\cat}\, D_{213}\right] 
\right. \nonumber \\ 
&& +m_t^2\, ({\cu}^2+\rho^2-8{\cu})\, \left[ {\cs}C_{12}+{\cu}_1(C_{23}+C_{14})-m_t^2\, {\cs}{\cu}\, 
D_{123}\right] -m_t^2\, ({\cat}^2+{\cu}^2-2\rho^2) \times \nonumber \\
&& \left.
({\cat}+{\cu}-8)\, C_{34} -2 m_t^4\, ({\cat}+{\cu}-8)\, ({\cat}{\cu}-\rho^2)\, (D_{123}+D_{213}+D_{132})\right\}\,.
\label{eq:FG}
\end{eqnarray}
The coefficients $C_\Delta$ and $C_\Box$ are given by
\begin{equation}
C_\Delta = \frac{3 m_h^2}{s-m^2_h}\, g^h_t\,,\ \ 
C_\Box = \left(g^h_{t}\right)^2\,,
\end{equation}
where $g_t^h \equiv \sqrt{2} m_t/v$ is the SM Yukawa coupling for the top quark.
The cross-section is calculated using {\tt FeynRules 2.0} \cite{Alloul:2013bka} and 
{\tt MadGraph5\_aMC@NLO} v2.3.3 \cite{Alwall:2014hca}.

\subsection{xSM4: A fourth chiral fermion generation in a type-II 2HDM}
%
Here we will summarize the essential points of Ref.\ \cite{Das:2017mnu}. One introduces a fourth chiral generation 
of fermions,
\beq
Q' = \begin{pmatrix}
	t' \\ b'
\end{pmatrix}\,,
L' = \begin{pmatrix}
	\nu'\\ \tau'
\end{pmatrix}\,,
\eeq
where all the components are massive enough to escape direct detection at the LHC and the masses are so arranged 
as to satisfy the bounds on both the oblique parameters $S$ and $T$. Note that a full chiral generation, if 
completely degenerate, gives a constant contribution of $2/3\pi$ to the oblique  $S$- parameter \cite{Kribs:2007nz,Dighe:2012dz}. To cancel this large contribution, one has to introduce a mass
splitting between the members of a doublet, and the splitting should go in opposite directions for the lepton and the 
quark doublets, to be in conformity with the $T$ parameter. Keeping this in mind, we will display our results 
for two benchmark points that satisfy the bounds coming from the oblique parameters.

Within the ambit of the SM, there is no way to play with the sign of the Yukawa couplings; they must have the 
same sign as the masses. This is also necessary to maintain the tree-level unitarity\cite{Bhattacharyya:2012tj}. However, if the scalar sector 
is extended to include another doublet, the wrong-sign limit, as shown below, may be achieved. 

Let us focus on a type-II 2HDM with a CP-conserving scalar potential, 
where the two doublets are denoted by $\Phi_1$ and $\Phi_2$. We denote the VEVs of the two doublets 
by $v_1/\sqrt{2}$ and $v_2/\sqrt{2}$ respectively, with $v=\sqrt{v_1^2+v_2^2}$
being the total electroweak VEV. The scalar spectrum has two CP-even neutral fields $h$ and $H$, with $h$ being the lighter one, one CP-odd
neutral field $A$ and a pair of charged fields $H^\pm$.  The physical fields 
$h$ and $H$ are obtained from the corresponding CP-even components of $\Phi_1$ and $\Phi_2$ by a rotation 
parametrized by the angle $\alpha$. The fact that the lighter CP-even scalar, $h$, has couplings with the SM gauge
bosons and fermions that are completely in conformity with the SM (within the experimental uncertainties) constrains
the allowed parameter space to lie close to the alignment limit, defined by,\cite{Gunion:2002zf,Carena:2013ooa,Bhattacharyya:2013rya,Bhattacharyya:2014oka,Bhattacharyya:2015nca}
\beq
\cos(\beta - \alpha) \approx 0\,.
\label{eq:align}
\eeq
On the other hand, to cancel the fourth generation contributions completely from the $gg\to h$, $h\to 
\gamma\gamma$, and $h\to Z\gamma$ processes, one needs the following correlation among the scaling of the Higgs couplings relative to the SM:
\beq
\frac{g_{VVh}}{g_{VVh}^{\rm SM}} \ = \ \frac{g_{uuh}}{g_{uuh}^{\rm SM}} \  = \ -\, \frac{g_{ddh}}{g_{ddh}^{\rm SM}} 
\ = \ -\, \frac{g_{\ell\ell h}}{g_{\ell\ell h}^{\rm SM}} \ = \ 1 \,,
\eeq
where $V$, $u$, $d$ and $\ell$ denote, generically, the weak gauge bosons, the up-type quarks, the down-type 
quarks and the charged leptons respectively. Such a `wrong-sign limit' can be obtained,
within the framework of a type-II 2HDM, as
\cite{Fontes:2014tga,Ferreira:2014dya,Biswas:2015zgk,Ferreira:2014naa}
\beq
\cos(\beta-\alpha) = 2\cot\beta ,\ \ {\rm with} \ \tan\beta\gg 2\,.
\label{eq:wsl}
\eeq
This makes the down-type Yukawa couplings with the nonstandard neutral scalar large (they are enhanced by 
$\tan\beta$) and for $b'$, the coupling can easily 
be nonperturbative. While this is an important issue that has recently been emphasized in Ref.\ \cite{Kang:2018jem}, 
we will keep our analysis confined to not-too-large values of $\tan\beta$, with a tacit understanding that 
any other issues with the stability of xSM4 are somehow taken care of, most probably by some other dynamics.

\section{Double Higgs production in xSM4}

Now that we have outlined how, in the wrong-sign limit, the type-II 2HDM can hide a chiral fourth 
generation in processes like $gg\to h$, $h\to\gamma\gamma$, $h\to Z \gamma$, let us concentrate 
on DHP: $gg \to hh$. The relevant Feynman diagrams can be generalized from those shown in Fig.\
\ref{f:SMfeynman}, by replacing $t$ with $t/b'/t'$ and the scalar propagator with $h/H$. We follow 
the same path as outlined in Section \ref{s:dhpsm}, except that (i) all the three heavy quarks, {\em viz.}, 
$t$, $t'$, and $b'$ will be taken into account\footnote{We refrain ourselves from going to such high
	values of $\tan\beta$ where the $b$-contributions become relevant and the Yukawa coupling for $b'$ 
	becomes badly nonperturbative.}, 
and (ii) apart from the $h$-mediated $\Delta$ diagrams, the $H$-mediated diagrams have 
to be considered too.

It is quite obvious that the cancellation of the 4th generation contributions due to the wrong-sign 
dynamics will no longer occur for the box diagrams, as the relevant Yukawa couplings appear 
twice in the box diagrams. Thus, we expect a large enhancement of the DHP rate over the SM prediction. Before we go 
into that, the $\Delta$-diagrams merit a comment. Note that there is a second $\Delta$ diagram, even in the absence 
of the 4th generation, that 
is mediated by the heavy scalar $H$. The $Hhh$ coupling plays an important role in the DHP process, as we 
will see later. However, in the exact alignment limit $\cos(\beta-\alpha) = 0$, the $Hhh$ coupling vanishes.

With more such $\Box$ and $\Delta$ diagrams in xSM4, expressions for whose amplitudes are 
analogous to what we have shown before, the
differential cross-section for DHP can be written as 
\begin{equation}
\frac{d\sigma_{gg\to hh} }{dt} = \frac{G^2_{F}\alpha^2_{s}} {256\, (2\pi)^3}\, 
\left[ \left\vert \sum_{t,t',b'} \{ (C^h_{\Delta}+ C^H_\Delta) 
F_{\Delta} + C_{\Box}F_{\Box}\} \right\vert^2 + \left\vert
\sum_{t,t',b'} C_{\Box} G_{\Box} \right\vert^2\right]\,.
\end{equation}
While $F_\Box$, $F_\Delta$, and $G_\Box$ have expressions analogous to those in Eq.\ 
(\ref{eq:FG}), with suitable replacement of the quark label, the expressions for $C_\Delta$ and 
$C_\Box$ are as follows:
\begin{equation}
C^h_\Delta = {\lambda}_{hhh}\, \frac{v}{s-m^2_h}\, g^h_q\,,\ \ \ 
C^H_\Delta = {\lambda}_{hhH}\, \frac{v}{s-m^2_H}\, g^H_q\,,\ \ \ 
C_{\Box} = \left(g^h_{q}\right)^2\,,
\end{equation}
where $g^{h/H}_q$ ($q = t, t', b'$) are given by
\begin{eqnarray}
&& g^h_q = \frac{\sqrt{2} m_q}{v}\, \frac{\cos\alpha}{\sin\beta}~~{\rm for}~q = t, t'\,,\ \ \ 
g^h_q = -\frac{\sqrt{2} m_q}{v}\, \frac{\sin\alpha}{\cos\beta}~~{\rm for}~q=b'\,, \nonumber\\
&& g^H_q = \frac{\sqrt{2} m_q}{v}\, \frac{\sin\alpha}{\sin\beta}~~{\rm for}~q = t, t'\,,\ \ \ 
g^H_q = \frac{\sqrt{2} m_q}{v}\, \frac{\cos\alpha}{\cos\beta}~~{\rm for}~q = b'\,,
\end{eqnarray}
and
\begin{eqnarray}
{\lambda}_{hhh} &=& -\frac{3}{v \sin2\beta}\, \left[ \lambda_S \, v^2 \cos(\alpha+\beta)\cos^2(\beta-\alpha) - m^2_{h}\, \{2\cos(\alpha+\beta)
+\sin 2\alpha\, \sin(\beta-\alpha)\} \right]\,, \\
{\lambda}_{hhH} &=& \frac{\cos(\beta-\alpha)}{v \sin2\beta}\, 
\left[ \sin 2\alpha\, \left(2m^2_{h} + m^2_{H}\right) - \frac{1}{2} \lambda_S \, v^2
(3\sin 2\alpha - \sin 2\beta)\right]\,,
\label{e:lHhh}
\end{eqnarray}
where the dimensionless quantity\cite{Das:2015mwa,Das:2015qva},
\begin{eqnarray}
	\lambda_S = \frac{2}{v^2} \frac{m_{12}^2}{\sin\beta\cos\beta} \,,
\end{eqnarray}
is used as a convenient parametrization for the soft $\mathbb{Z}_2$ breaking effect
in the scalar potential.
 For our analysis, we will use the dimensionless coupling $\tilde{g}$, 
defined as
\begin{equation}
\tilde{g}_{hhh(H)} \equiv \frac{v}{3m_h^2}\, \lambda_{hhh(H)}\,.
\end{equation}

Note that, once the wrong-sign limit of Eq.~(\ref{eq:wsl}) is imposed and some benchmark values
for the nonstandard masses are chosen, the trilinear couplings are controlled by $\tan\beta$
and $\lambda_S$. Therefore, $\lambda_S$ can be tuned properly to adjust the value of
$\tilde{g}_{hhH}$ and consequently, the DHP cross section can be treated as a function of
$\tilde{g}_{hhH}$ and $\tan\beta$.

We will start with the limit $\tilde{g}_{hhH}=0$, {\it i.e.}, when the $H$-mediated diagram
is absent. In this case, the box amplitude is expected to pick up a factor of 3 compared to that
in the SM due to the presence of three heavy quarks, $t$, $t'$ and $b'$. Also notice that the
$t'$ and $b'$ contributions should cancel each other in the $h$ mediated triangle amplitudes because
of the wrong-sign limit. Since the $t$ mediated triangle amplitude is also subdominant, this will
lead to an enhancement of the DHP cross-section approximately by a factor of 9, which can be
sensed in the near future.

Next we slowly switch on $\tilde{g}_{hhH}$ till the experimental upper bound on the DHP cross-section 
$\approx 330-340$ fb \cite{Aaboud:2018sfw} is reached.  Thus, for a given $\tan\beta$, one obtains
an upper bound on $\vert \tilde{g}_{hhH}\vert$, although the exact number depends on the sign of this 
coupling. We display our results for two different benchmark points as shown in Table~\ref{tab:bm}. 
The masses specify all the relevant parameters, like the scaled Yukawa couplings $g^{h(H)}_q$, 
as $\tan\beta$ is treated as a variable parameter and the mixing angle $\alpha$ is obtained from 
Eq.\ (\ref{eq:wsl}). The soft breaking term $\lambda_S$ is adjusted so that that the DHP cross-section 
reaches the experimental limit at $\tan\beta=20$. 
The masses of the heavy scalars and the fourth generation fermions are chosen in such 
a way as to evade the direct detection limits \cite{Chatrchyan:2012fp,Aad:2015tba}.
We have explicitly checked that the oblique parameters $S$ and $T$ for both the benchmark points as shown 
in Table \ref{tab:bm}, coming from fermionic \cite{Kribs:2007nz,Dighe:2012dz} and scalar \cite{Grimus:2007if,
	Grimus:2008nb} loops, 
are within the experimental limits given by \cite{Tanabashi:2018oca}
\begin{equation}
\Delta S = 0.05\pm 0.10 \,, \ \ \  \Delta T = 0.08\pm 0.12 \,.
\label{e:stlims}
\end{equation}
 While computing the $H$-mediated $\Delta$ diagrams, we have taken into account the finite, possibly 
 non-negligible widths of the nonstandard scalars in our numerical codes. However, the $H$-mediated 
 diagrams are not the primary reasons behind the enhancement in the DHP cross section. Therefore our 
 main result will not crucially depend on the effects arising due to the finite width of the nonstandard 
 scalars.

\begin{table}[htbp!]
	\begin{center}
		\begin{tabular}{ |c|c|c|c|c|c|c|c|c| } 
			\hline
			Benchmark & $m_{t'}$ & $m_{b'}$& $m_{\tau'}$ & $m_{\nu'}$ & $m_H$ & $m_H^{\pm}$ & $m_A$  \\ 
			\hline
			BP1 &1430 & 1380 & 1380 & 495  &   1010   &    1900  &   2800 \\
			\hline
			BP2 &1430 & 1380 & 1380 & 500  &   2160    &    2650   &   4050  \\
			\hline
		\end{tabular}
		\caption{\em The benchmark points BP1 and BP2, with all masses in GeV.}		
		\label{tab:bm}
	\end{center}
\end{table}

\section{Observability of xSM4}
As explained in the previous section,
in the wrong-sign limit, $\tilde{g}_{hhH}$ and $\tan\beta$ are the only free parameters relevant for 
our study once we decide to stick to the chosen benchmark points. Keeping in mind that the up and down 
type Yukawa couplings in the 4th generation do not become badly non-perturbative and at the same 
time to be consistent with the wrong-sign limit, we confine ourselves to the range $3 < \tan\beta < 20$.

\begin{figure}[htbp!]
	\centering
	%
	\includegraphics[width=0.60\textwidth]{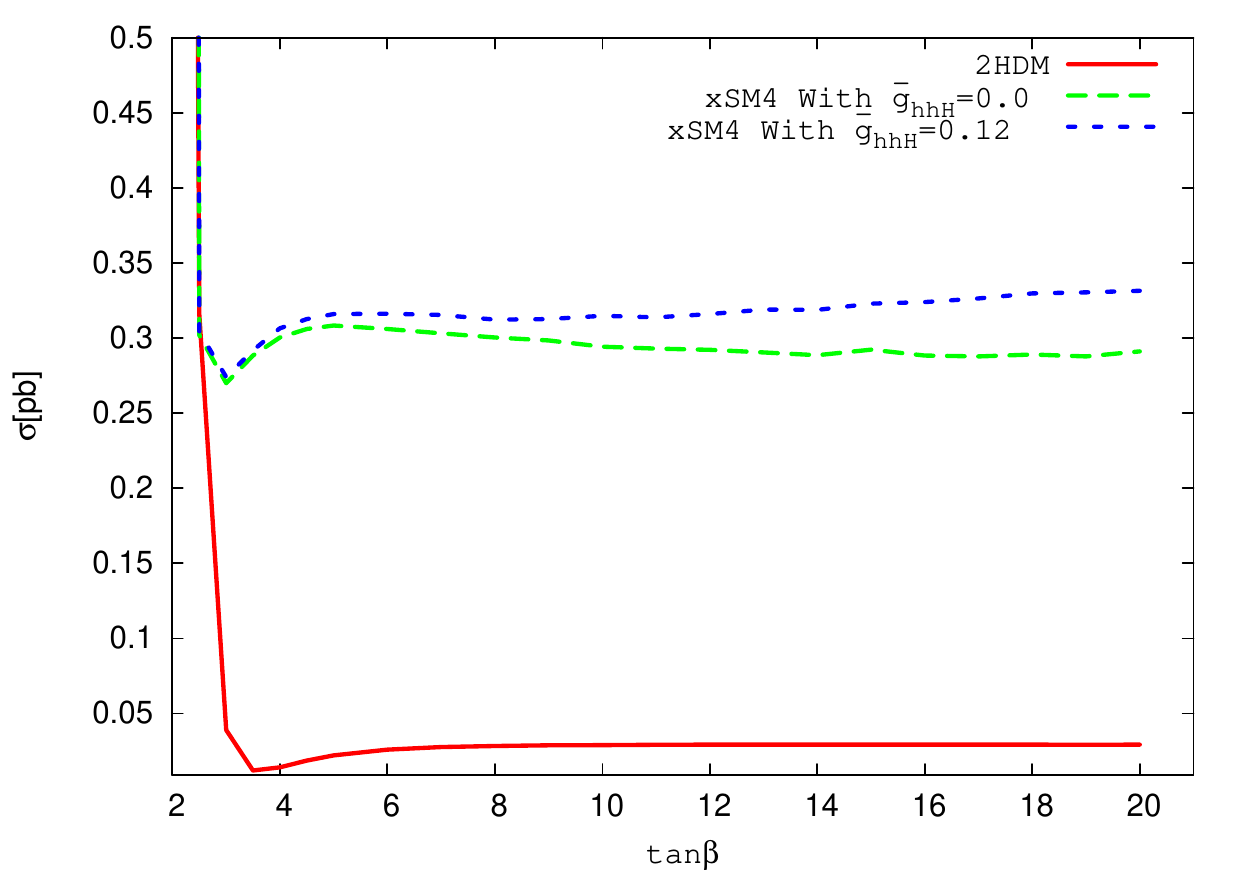}
	\caption{Double Higgs production cross-section as a function of $\tan\beta$ for the
		Benchmark point 1.}
	\label{gg2hh_mhB1}
\end{figure}

First, let us focus on the case where the 4th generation is present but the $H$-mediated diagrams are
absent, possibly because of a vanishing $\tilde{g}_{hhH}$. The $t'$ and $b'$ mediated Box amplitudes, on 
top of the $t$ mediated one, enhances the SM DHP cross-section approximately by a factor of 9, as can be seen 
from Fig.\ \ref{gg2hh_mhB1}, to about 300 fb. 
This is in contrast to the enhancement of the DHP cross-section \cite{Basler:2018dac} due to resonant production of heavy
scalars and their subsequent decays.
 The nature of the plot near $\tan\beta=3$ can be explained
by the imperfect cancellation of the $t'$ and $b'$ amplitudes in the $h$-mediated triangle diagrams.
However, the cross-section stabilizes for moderate or large values of $\tan\beta$, {\it i.e.}, when
we are very close to the wrong-sign limit.
We have checked that the nature of the plot remains 
the same for both the benchmark points. The region near $\tan\beta = 3$ is very close to, but not, 
the actual minimum of DHP cross-section in xSM4. This will be discussed later in this Section.

The higgs pair going to bb$\tau$$\tau$ channel also studied in \cite{Dolan:2012rv}.
Among the all the probable final states to which $h$ can decay into, $h \, h \to b\bar{b}\tau^+\tau^-$ appears 
to be one of the most promising channels, because of a relatively small background compared to other final 
states \cite{CMS:2017orf}. The branching 
fraction of $hh\to b\bar{b} \tau^+ \tau^-$ is about $7.3\%$ \cite{DiMicco:2019ngk} with the $b$-tagging efficiency 
being about $75\%$, and the $\tau$-tagging efficiency being about $50\%$ \cite{Bagliesi:2007qx,Tanasijczuk:2013sya,Katz:2010iq}. So the number of events 
observed, with an integrated luminosity of 300 fb$^{-1}$, is about 1200, which should be
 more than ample to verify the existence of the fourth generation in the wrong sign limit\footnote{The present upper limit in this 
channel is $12.7$ and $31.4$ times that of the SM, 
	by ATLAS and CMS Collaborations respectively.}.

\begin{figure}[htbp!]
	\centering
	\includegraphics[width=0.52\textwidth]{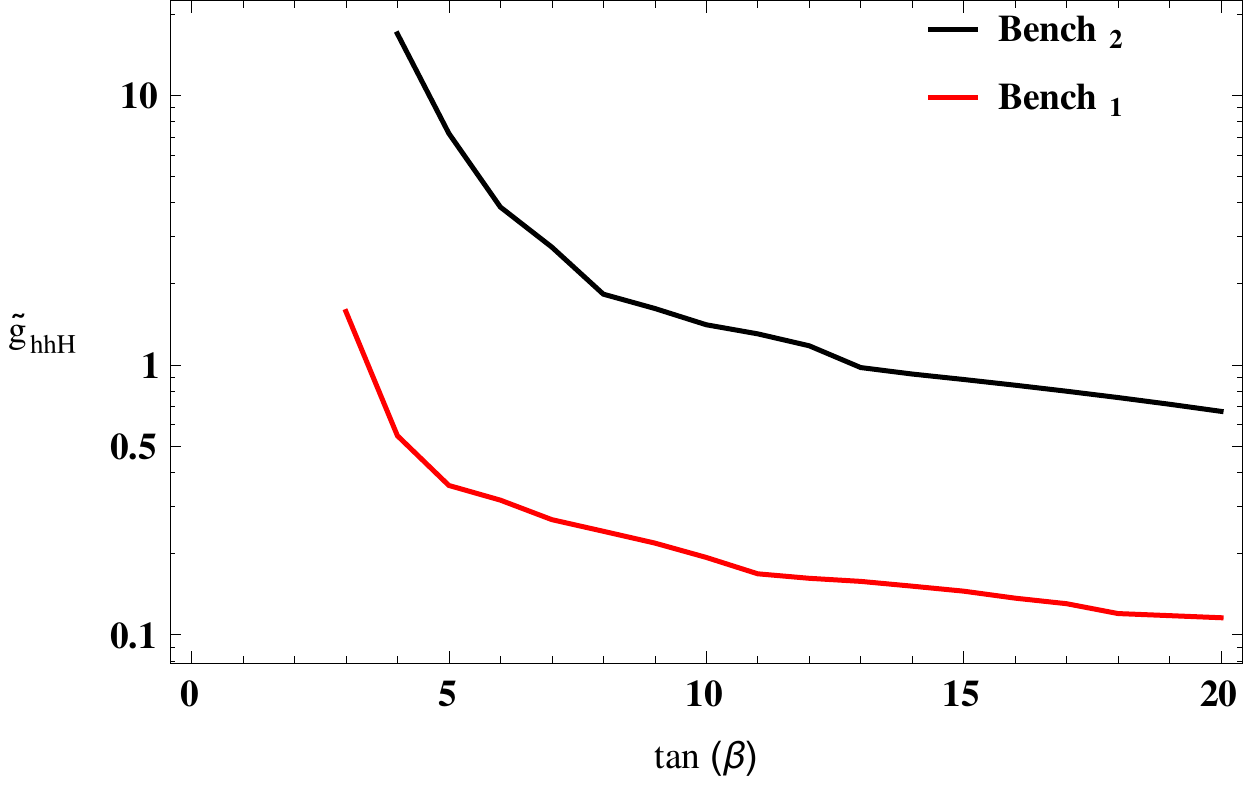}
	\caption{
	The upper limit of $\tilde{g}_{hhH}$ as a function of $\tan \beta$ for the two benchmark points, with 
	$M_H = 1.01$ TeV and 2.16 TeV respectively. Regions above the respective lines are ruled out from the 
	experimental limit on DHP.}
	
		\label{fig:ghhH}	
\end{figure}

The upper limit of the DHP cross-section, {\em viz.}, 330-340 fb \cite{Aaboud:2018sfw}, is still 
 somewhat above the enhancement caused by the fourth generation alone. This gap may be 
bridged with the contribution coming from the $H$-mediated $\Delta$ diagrams. This, in turn, sets an 
upper limit on $\tilde{g}_{hhH}$ as a function of $\tan\beta$, as displayed in Fig.\ \ref{fig:ghhH}. 
The limits are shown for the two benchmark points BP1 and BP2, with $m_H=1.01$ and $2.16$ TeV
respectively. The limits become stronger with increasing $\tan\beta$, as the $\Delta$ diagram with $b'$ 
starts to become more significant.  Obviously, the $H$-propagator suppression ensures that the limit on
$\tilde{g}_{hhH}$ is higher for BP2 than for BP1; for $\tan\beta=20$, the coupling can be as large as 
$0.67$ for BP2 but only $0.12$ for BP1.

\begin{figure}[htbp!]
	\centering
	\includegraphics[width=0.52\textwidth]{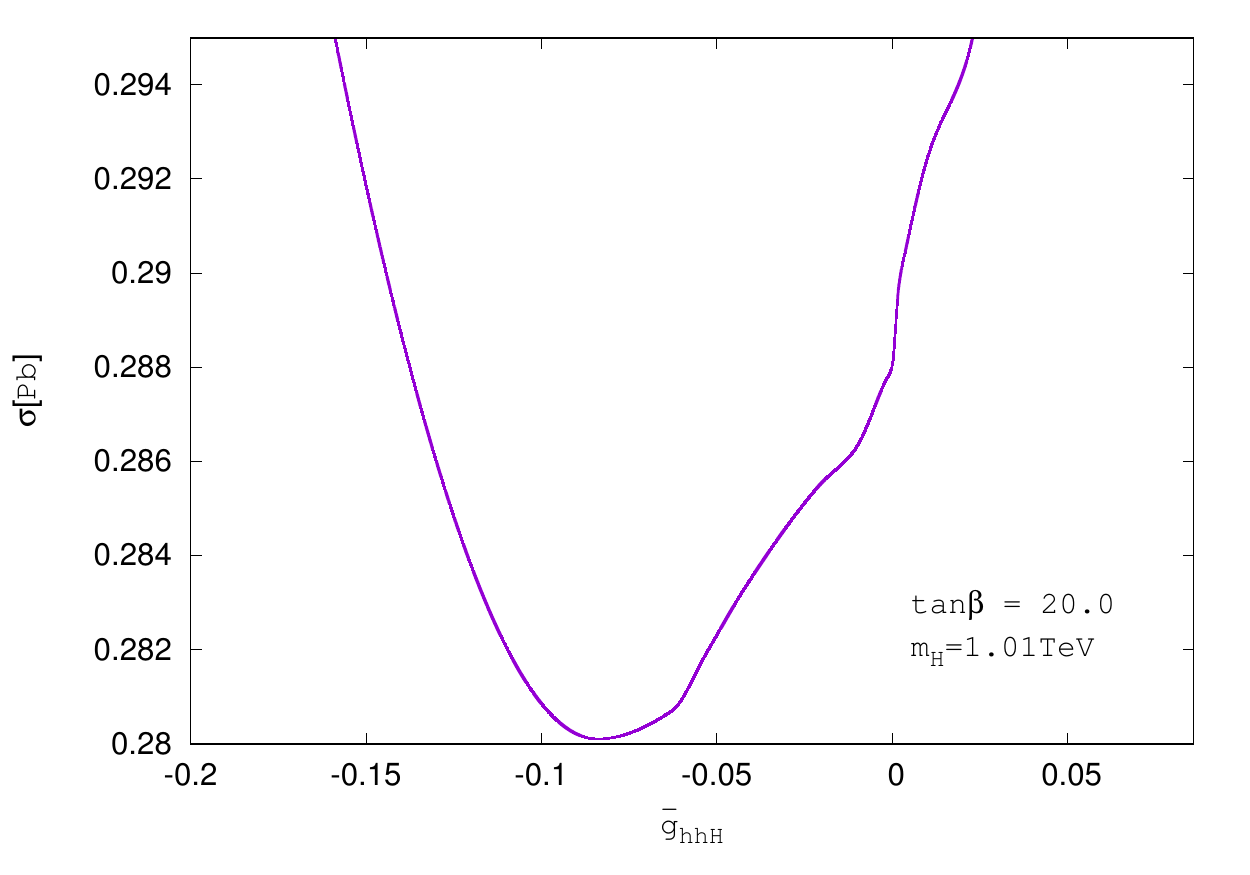}
	\caption{Variation of the DHP cross-section with $\tilde{g}_{Hhh}$, for 
	 $\tan \beta$=20.}
	\label{f:Interference}	
\end{figure}

The coupling $\tilde{g}_{hhH}$ can, in principle, be negative too. Note that the sign is important only for 
the interference terms in the squared amplitude. It is easy to check that there is a destructive interference 
between the $b'$-mediated $\Delta$ diagram with the other $\Box$ amplitudes, which slightly lowers the 
DHP cross-section from its $\tilde{g}_{hhH}=0$ value. This is shown in Fig.\ \ref{f:Interference}, where we 
have taken $\tan\beta=20$ to maximise the decrease. The lowest point of the curve is 
actually the minimum possible DHP cross-section ($\sim 280$ fb) in xSM4, not the one that one gets 
from the absence of the $\Delta$ diagrams. However, the difference is negligible compared to the 
theoretical uncertainties in the estimation of the cross-section as well as the experimental 
uncertainties, 
and even more so when the higher-order effects are neglected, as mentioned in the Introduction.

\section{Conclusion}

It is well-known that DHP is the only process that can realistically probe the Higgs self-coupling $\lambda$. 
In a 2HDM that conserves CP, and that has all other scalars at 1 TeV or more except the 125 GeV Higgs, 
the DHP can probe the couplings $\lambda_{hhh}$ and $\lambda_{hhH}$, the latter in the large 
$\tan\beta$ limit that enhances the bottom quark coupling to $H$. 

In this paper, we show how DHP may be used to unveil a fourth chiral generation that can possibly hide in the 
LHC data. Such a model has been discussed in detail in Ref.\ \cite{Das:2017mnu}, which is embedded in a Type-II 2HDM. 
Even in the small parameter space where the fourth generation effects on Higgs production and decay cancel 
out, the double Higgs production can successfully probe the model. This is because no such cancellation mechanism 
works for DHP, and one expects about an order of magnitude enhancement over the SM prediction, 
which is close to the experimental limit. Thus, LHC at 14 TeV has an excellent chance to discover the signal,
or rule the model out.

DHP limits can also rule out a significant portion of the parameter space of the scalar sector of xSM4, with the 
constraints on $\lambda_{hhH}$ and  $\tan\beta$. Admittedly, this also depends on other 
parameters of the potential, so one needs to look for other corroborative signals. 

The analysis was performed at the leading order, and the higher-order effects may be 
extremely important, particularly when the Yukawa couplings are nonperturbative. Including such 
corrections should increase the DHP cross-section and hence tighten our limits, or in an extreme case, 
may rule out the model already. Even with this caveat, this remains an important and interesting channel to probe.

Before we conclude, it should be re-emphasized that probing the trilinear self-coupling of the Higgs
bosons has, so far, been the driving motivation behind the search for DHP at the LHC. Quite
evidently, we can add to the stimulus for di-Higgs searches by making this channel a sensitive tool
for unveiling certain new physics models. In this paper, we have done precisely that. We have shown
that, an extra sequential generation of fermions with wrong-sign Yukawa couplings, which can remain completely
hidden in single Higgs production and decay \cite{Das:2017mnu}, can potentially reveal themselves
exclusively in the di-Higgs searches. To our knowledge, this is the first time when such an
exclusivity for the DHP in the context of new physics searches has been pointed out explicitly.
Therefore, we hope that our current study will encourage our experimental colleagues to view
the di-Higgs searches in a new light.



{\em Ackmowledgements}: A.K. acknowledges the Science and Engineering Research Board, Government of India, 
for support.

\bibliographystyle{JHEP}
\bibliography{diHiggs-PRD.bib}

\end{document}